\documentclass[12pt]{article}
\usepackage{a4}
\usepackage{amsfonts}
\usepackage{amsmath,amssymb}
\usepackage{graphicx}

\usepackage{multirow}
\usepackage[margin=0.8in]{geometry}

\newcommand{\overbar}[1]{\mkern1.5mu\overline{\mkern-1.5mu#1\mkern-1.5mu}\mkern 1.5mu}

\makeatother

\def\un#1{\relax\ifmmode\@@underline#1\else
        $\@@underline{\hbox{#1}}$\relax\fi}

\let\du=\du                     % dot-under
\def\bo{{\raise-.3ex\hbox{\large$\Box$}}}               % D'Alembertian
\def\TH{{\raise.2ex\hbox{$\displaystyle \bigodot$}\mskip-4.7mu \llap H \;}}
\def\face{{\raise.2ex\hbox{$\displaystyle \bigodot$}\mskip-2.2mu \llap {$\ddot
        \smile$}}}                                      % happy face
\def\leftrightarrowfill{$\mathsurround=0pt \mathord\leftarrow \mkern-6mu
        \cleaders\hbox{$\mkern-2mu \mathord- \mkern-2mu$}\hfill
        \mkern-6mu \mathord\rightarrow$}
\def\dvec#1{\vbox{\ialign{##\crcr
        \leftrightarrowfill\crcr\noalign{\kern-1pt\nointerlineskip}
        $\hfil\displaystyle{#1}\hfil$\crcr}}}           % <--> accent
\def\frac#1#2{{\textstyle{#1\over\vphantom2\smash{\raise.20ex
        \hbox{$\scriptstyle{#2}$}}}}}                   % fraction
\def\sfrac#1#2{{\vphantom1\smash{\lower.5ex\hbox{\small$#1$}}\over
        \vphantom1\smash{\raise.4ex\hbox{\small$#2$}}}} % alternate fraction
\def\bfrac#1#2{{\vphantom1\smash{\lower.5ex\hbox{$#1$}}\over
        \vphantom1\smash{\raise.3ex\hbox{$#2$}}}}       % "
\def\afrac#1#2{{\vphantom1\smash{\lower.5ex\hbox{$#1$}}\over#2}}    % "
\def\[{\lfloor{\hskip 0.35pt}\!\!\!\lceil}
\def\]{\rfloor{\hskip 0.35pt}\!\!\!\rceil}

\def\du#1#2{_{#1}{}^{#2}}

\def\un{\underline}
\def\fracmm#1#2{{{#1}\over{#2}}}

\def\low#1{{\raise -3pt\hbox{${\hskip 0.75pt}\!_{#1}$}}}

\newskip\humongous \humongous=0pt plus 1000pt minus 1000pt

\newif\ifdtup

\newcommand{\be}{\begin{equation}}
\newcommand{\ee}{\end{equation}}
\newcommand{\nbe}{\begin{equation*}}
\newcommand{\nee}{\end{equation*}}

\begin{document}

\thispagestyle{empty}

{\hbox to\hsize{
\vbox{\noindent March 2017 \hfill IPMU17-0007 }}}
{\hbox to\hsize{
\vbox{\noindent  revised version \hfill }}}
\noindent
\vskip2.0cm
\begin{center}

{\Large\bf Higgs mechanism and cosmological constant 
\vglue.1in
in $N=1$ supergravity with inflaton in a vector multiplet
}

\vglue.3in

Yermek Aldabergenov~${}^{a}$  and Sergei V. Ketov~${}^{a,b,c}$ 
\vglue.1in

${}^a$~Department of Physics, Tokyo Metropolitan University, \\
Minami-ohsawa 1-1, Hachioji-shi, Tokyo 192-0397, Japan \\
${}^b$~Kavli Institute for the Physics and Mathematics of the Universe (IPMU),
\\The University of Tokyo, Chiba 277-8568, Japan \\
${}^c$~Institute of Physics and Technology, Tomsk Polytechnic University,\\
30 Lenin Ave., Tomsk 634050, Russian Federation \\
\vglue.1in
aldabergenov-yermek@ed.tmu.ac.jp, ketov@tmu.ac.jp
\end{center}

\vglue.3in

\begin{center}
{\Large\bf Abstract}
\end{center}
\vglue.1in
\noindent  The $N=1$ supergravity models of cosmological inflation with inflaton belonging to a massive vector multiplet
and spontaneous SUSY breaking after inflation are reformulated as the supersymmetric $U(1)$ gauge theories of a massless
vector superfield interacting with the Higgs and Polonyi chiral superfields, all coupled to supergravity. The $U(1)$ gauge sector
is identified with the $U(1)$ gauge fields of the super-GUT coupled to supergravity, whose gauge group has a $U(1)$ factor.
A positive cosmological constant (dark energy) is included. The scalar potential is calculated, and its de Sitter vacuum solution
is found to be stable.
\newpage

\section{Introduction}

PLANCK observations \cite{Ade:2015xua, Ade:2015lrj,Array:2015xqh} of the Cosmic Microwave Background (CMB) radiation 
favour chaotic slow-roll inflation in its single-field realization,  i.e. the large-field inflation driven by a single scalar called {\it inflaton} with an approximately flat scalar potential.

Embedding inflationary models into $N=1$ four-dimensional supergravity is needed to connect them to particle physics 
theory beyond the Standard Model of elementary particles and to quantum gravity.
  Most of the literature about inflation in supergravity is based on an assumption that the inflaton belongs to a chiral (scalar) multiplet --- see e.g., the reviews \cite{Yamaguchi:2011kg,Ketov:2012yz}. However, the inflaton can also
be assigned to a massive $N=1$ {\it vector} multiplet. It has some theoretical advantages because there is only one real scalar in an $N=1$ massive vector multiplet. The $\eta$-problem does not arise because the scalar potential of a vector multiplet in supergravity is of the $D$-type instead of the $F$-type. The minimal inflationary models with the inflaton belonging to a massive vector multiplet were constructed in 
Ref.~\cite{Ferrara:2013rsa} by exploiting the non-minimal self-coupling of a vector multiplet to supergravity \cite{VanProeyen:1979ks}. The supergravity inflationary models \cite{Ferrara:2013rsa} have the single-field scalar potential given by an
arbitrary real function squared. Those scalar potentials are always bounded from below and allow any
desired values of the CMB observables $n_s$ and $r$. However, the minima of the scalar potentials of \cite{Ferrara:2013rsa} 
have a vanishing cosmological constant and the vanishing Vacuum Expectation Value (VEV) of the auxiliary field $D$, so that they allow only  Minkowski vacua where supersymmetry is restored after inflation. 

A simple extension of the inflationary models \cite{Ferrara:2013rsa} was proposed in Ref.~\cite{Aldabergenov:2016dcu}  by adding 
a {\it Polonyi} (chiral) multiplet \cite{Polonyi:1977pj} with a linear superpotential.  The inflationary models \cite{Aldabergenov:2016dcu} also accommodate arbitrary values of  $n_s$ and $r$, and have a Minkowski vacuum after inflation, but {\it with} spontaneously broken supersymmetry (SUSY). In this paper we further extend the models of Ref.~\cite{Aldabergenov:2016dcu}  by allowing them to have a
{\it positive} cosmological constant, i.e. a {\it de-Sitter} vacuum after inflation.

Yet another motivation comes from an exposition of the super-Higgs effect in supergravity by presenting the new $U(1)$ gauge-invariant form of the class of inflationary models under investigation. This paves the way towards embedding our models into the superymmetric Grand Unification Theories (sGUT) coupled to supergravity, when they have a spontaneously broken $U(1)$ factor in the sGUT gauge group. The
physical scale of cosmological inflation can be identified with the Hubble (curvature) scale $H\approx 10^{14}$ GeV or the
inflaton mass $m_{\rm inf}\approx 10^{13}$ GeV. The inflationary scale is thus less (though, not much less!) than   the sGUT scale of  $10^{16}$ GeV. The simple sGUT groups $SU(5)$, $SO(10)$ and $E_6$ are well motivated in the Calabi-Yau compactified heterotic strings, however, they usually come with at least one extra "undesired" $U(1)$ factor in the gauge group. The well known examples include the gauge symmetry breaking $E_6\to SO(10)\times U(1)$,  $SO(10)\to SU(5)\times U(1)$, and the "flipped" $SU(5)\times U_X(1)$ sGUT originating from heterotic strings. Exploiting the Higgs mechanism in supergravity allows us to propose an identification of the $U(1)$ gauge vector multiplet of those sGUT models with the  inflaton vector multiplet we consider, thus unifying inflation with those sGUT in supergravity. Besides the sGUT gauge unification, related proton decay and baryon number violation, having the $U(1)$ factor in the sGUT gauge group allows one to get rid of monopoles, because the gauge group is not semi-simple \cite{Ketov:1996bm}. And having a positive cosmological constant  takes into account dark energy too.
 
Our paper is organized as follows. In Sec.~2 we briefly review the supergravity models \cite{Aldabergenov:2016dcu}. In Sect.~3 we present their $U(1)$ gauge-invariant formulation and the Higgs mechanism. A positive cosmological constant is added in Sec.~4. The scalar potential and it stability are studied in Sec.~5.  Our conclusion is given by Sec.~6.

\section{Scalar potential and SUSY breaking with a massive vector multiplet in the absence of a cosmological constant}

The inflationary models of Ref.~\cite{Aldabergenov:2016dcu} are defined in  curved superspace of $N=1$ supergravity 
 \cite{Wess:1992cp} by the Lagrangian ($M_{\rm Pl}=1$)~\footnote{Our notation and conventions coincide with the standard ones in Ref.~\cite{Wess:1992cp}, including the spacetime signature $(-,+,+,+)$. The $N=1$ superconformal calculus \cite{Ferrara:2013rsa,VanProeyen:1979ks} after the superconformal gauge fixing is equivalent to the curved superspace description of $N=1$ Poincar\'e supergravity.} 
\begin{equation} \label{sslag}
\mathcal{L}=\int d^2\theta 2\mathcal{E}\left\lbrace \frac{3}{8}(\overbar{\mathcal{D}}\overbar{\mathcal{D}}-8\mathcal{R})e^{-\frac{1}{3}(K+2J)}+\frac{1}{4}W^\alpha W_\alpha +\mathcal{W} \right\rbrace +{\rm h.c.}~,
\end{equation}
in terms of chiral superfields $\Phi_i$, representing ordinary (other than inflaton) matter, with a K{\"a}hler potential $K=K(\Phi_i,\overbar{\Phi}_i)$ and a chiral superpotential $\mathcal{W}=\mathcal{W}(\Phi_i)$, and interacting with the vector (inflaton) superfield $V$ described by a real function $J=J(V)$ and having the superfield strength $W_\alpha\equiv-\frac{1}{4}(\overbar{\mathcal{D}}\overbar{\mathcal{D}}-8\mathcal{R})\mathcal{D}_\alpha V$. We have also introduced the chiral density superfield $2\mathcal{E}$ and the chiral scalar curvature superfield $\mathcal{R}$ \cite{Wess:1992cp}.

After eliminating the auxiliary fields and changing the initial (Jordan) frame to Einstein frame, the bosonic part of the Lagrangian 
(\ref{sslag}) reads \cite{Aldabergenov:2016dcu} 
\begin{equation} \label{complag}
e^{-1}\mathcal{L}=-\frac{1}{2}R-K_{ij*}\partial_m A_i\partial^m\bar{A}_j-\frac{1}{4}F_{mn}F^{mn}-\frac{1}{2}J''\partial_mC\partial^mC-\frac{1}{2}J''B_mB^m-\mathcal{V}~,
\end{equation}
and has the scalar potential
\begin{equation} \label{pot}
\mathcal{V}=\frac{1}{2}{J'}^2+e^{K+2J}{}\biggl[
	K^{-1}_{ij^*}(\mathcal{W}_i+K_i\mathcal{W})(\overbar{\mathcal{W}}_j+K_{j^*}\overbar{\mathcal{W}})-\bigg(3-2\fracmm{{J'}^2}{J''}\bigg)\mathcal{W}\overbar{\mathcal{W}}
	\biggl]~,
\end{equation}
where we have introduced the vierbein determinant $e\equiv\text{det} e_m^a$, the spacetime scalar curvature $R$, the complex scalars $A_i$ as physical components of $\Phi_i$; the real scalar $C$ and the real vector $B_m$, with the corresponding field strength $F_{mn}=\mathcal{D}_mB_n-\mathcal{D}_nB_m$, as physical components of $V$. The functions $K$, $J$ and $\mathcal{W}$ now represent the lowest components ($A_i$ and $C$) of the corresponding  superfields. As regards their derivatives, we use the notation $K_i\equiv\frac{\partial K}{\partial A_i}$, $K_{i^*}\equiv\frac{\partial K}{\partial\overbar{A}_i}$, $K_{ij^*}\equiv\frac{\partial^2K}{\partial A_i\partial\overbar{A}_j}$, $J'\equiv\frac{\partial J}{\partial C}$, $\mathcal{W}_i\equiv\frac{\partial\mathcal{W}}{\partial A_i}$, $\overbar{\mathcal{W}}_i\equiv\frac{\partial\overbar{\mathcal{W}}}{\partial\overbar{A}_i}$. As is clear from Eq.~\eqref{complag}, the absence of ghosts requires $J''(C)>0$, where the primes denote differentiations with respect to the given argument.~\footnote{Our $J$-function differs by the sign from that in Ref.~\cite{Ferrara:2013rsa,VanProeyen:1979ks}.}

For our purposes here, we restrict ourselves to a single chiral superfield $\Phi$ whose K\"ahler potential and the superpotential are those
of the {\it Polonyi model} \cite{Polonyi:1977pj}:
\begin{equation} \label{polonyi}
K= \Phi\overbar{\Phi}~,\qquad \mathcal{W}=\mu(\Phi +\beta)~,
\end{equation}
with the parameters $\mu$ and $\beta$. The choice (\ref{polonyi}) is quite natural (and unique) for a nilpotent (Volkov-Akulov)
chiral superfield $\Phi$ obeying the constraint $\Phi^2=0$, though we do not employ the nilpotency condition here, in order to avoid its possible clash with unitarity at high energies.

A substitution of Eq.~\eqref{polonyi} into the Lagrangian \eqref{complag} yields
\begin{multline}
e^{-1}\mathcal{L}=-\frac{1}{2}R-\partial_mA\partial^m\bar{A}-\frac{1}{4}F_{mn}F^{mn}-\frac{1}{2}J''\partial_mC\partial^mC-\frac{1}{2}J''B_mB^m-\frac{1}{2}{J'}^2\\-\mu^2e^{A\bar{A}+2J}\biggl[
	|1+A\beta+A\bar{A}|^2-\bigg(3-2\fracmm{{J'}^2}{J''}\bigg)|A+\beta|^2
	\biggl]~,
\end{multline} 
where the complex scalar $A$ is the lowest component of the Polonyi chiral superfield $\Phi$.

The Minkowski vacuum conditions
\begin{gather}
V=\frac{1}{2}{J'}^2+\mu^2e^{A\bar{A}+2J}\biggl[
	|1+A\beta+A\bar{A}|^2-\bigg(3-2\fracmm{{J'}^2}{J''}\bigg)|A+\beta|^2
	\biggl]=0~,\label{vac1}\\
\partial_{\bar{A}}V=\mu^2e^{A\bar{A}+2J}\biggl[
A(1+\bar{A}\beta+A\bar{A})+(A+\beta)(1+A\beta+A\bar{A})-\bigg(3-2\fracmm{{J'}^2}{J''}\bigg)(A+\beta)\nonumber\\
+A|1+A\beta+A\bar{A}|^2-\bigg(3-2\fracmm{{J'}^2}{J''}\bigg)A|A+\beta|^2\biggl]=0~,\label{vac2}\\
\partial_C V=J'\biggl\{J''+2\mu^2e^{A\bar{A}+2J}\biggl[
	|1+A\beta+A\bar{A}|^2-\bigg(1-2\fracmm{{J'}^2}{J''}+\fracmm{J'J'''}{{J''}^2}\bigg)|A+\beta|^2
	\biggl]\biggl\}=0~,\label{vac3}
\end{gather}
can be satisfied when $J'=0$ that separates the Polonyi multiplet from the vector multiplet. The Polonyi field VEV is then given by $\langle A\rangle=(\sqrt{3}-1)$ and $\beta=2-\sqrt{3}$ \cite{Polonyi:1977pj}. This solution describes a {\it stable} Minkowski vacuum with spontaneous SUSY breaking at an arbitrary scale $\langle F\rangle=\mu$. The related gravitino mass (at the minimum having $J'=0$) is given by $m_{3/2}=\mu e^{2-\sqrt{3}}$. There is also a massive scalar of mass $2m_{3/2}$ and a massless fermion in the physical spectrum. 

As a result, the Polonyi field does not affect the inflation driven by the inflaton scalar $C$ belonging to the massive vector multiplet and having the $D$-type scalar potential $V(C)=\frac{1}{2}{J'}^2$ with a real $J$-function.  Of course, the true inflaton field should be canonically normalized via the proper field redefinition of $C$.

\section{Massless vector multiplet and super-Higgs mechanism}

The matter-coupled supergravity model (\ref{sslag}) can also be considered as a supersymmetric (Abelian, non-minimal) gauge theory (coupled to supergravity and a Higgs superfield) in the (supersymmetric) gauge where the Higgs superfield is gauged away (say,
equal to $1$). When the gauge $U(1)$ symmetry is restored by introducing back the Higgs (chiral) superfield, the vector superfield $V$ becomes the gauge superfield of a spontaneously broken $U(1)$ gauge group. In this Section we restore the gauge symmetry in the way consistent with local supersymmetry, and then compare our results with those of the previous Section.

We start with a Lagrangian having the same form as  (\ref{sslag}) ,
\begin{equation}
\mathcal{L}=\int d^2\theta 2\mathcal{E}\left\lbrace \frac{3}{8}(\overbar{\mathcal{D}}\overbar{\mathcal{D}}-8\mathcal{R})e^{-\frac{1}{3}(K+2J)}+\frac{1}{4}W^\alpha W_\alpha +\mathcal{W}(\Phi_i) \right\rbrace +{\rm h.c.}~,\label{hsslag}
\end{equation}
where $K= K(\Phi_i,\overbar{\Phi}_j)$ and the indices $i,j,k$ refer to the chiral (matter) superfields, {\it excluding} the Higgs chiral superfield that we denote as $H,\overbar{H}$. Now, in contrast to the previous Section, the real function $J$ also depends on the Higgs superfield as $J=J(He^{2V}\overbar{H})$, while the vector superfield $V$ is {\it massless}. The Lagrangian \eqref{hsslag} is invariant under the supersymmetric $U(1)$ gauge transformations
\begin{gather}
H\rightarrow H'=e^{-iZ}H~,\;\;\;\overbar{H}\rightarrow \overbar{H}'=e^{i\overbar{Z}}\overbar{H}~,\label{supergh}\\
V\rightarrow V'=V+\frac{i}{2}(Z-\overbar{Z})~,\label{supergv}
\end{gather}
whose gauge parameter $Z$ itself is a chiral superfield. The Lagrangian (\ref{sslag}) of Sec.~2 is recovered from Eq.~(\ref{hsslag})
in the gauge $H=1$, after the redefinition $J_{\rm new}(e^{2V})=J_{\rm old}(V)$.

The $U(1)$ gauge symmetry of the Lagrangian (\ref{hsslag}) allows us to choose a different ({\it Wess-Zumino}) 
supersymmertic gauge by "gauging away" the chiral and anti-chiral parts of the general superfield $V$ via the appropriate choice of the superfield parameters $Z$ and $\overbar{Z}$ as
\begin{gather*}\label{wzgauge}
V|=\mathcal{D}_\alpha\mathcal{D}_\beta V|=\overbar{\mathcal{D}}_{\dot{\alpha}}\overbar{\mathcal{D}}_{\dot{\beta}} V|=0,\\
\overbar{\mathcal{D}}_{\dot{\alpha}}\mathcal{D}_{\alpha}V|={\sigma_{\alpha\dot{\alpha}}}^m B_m~,
\\
\mathcal{D}_\alpha W^{\beta}|=\frac{1}{4}{\sigma_{\alpha\dot{\alpha}}}^m\overbar{\sigma}^{\dot{\alpha}\beta n}(2iF_{mn})+{\delta_\alpha}^\beta D~,
\\\overbar{\mathcal{D}}\overbar{\mathcal{D}}\mathcal{D}\mathcal{D}V|=\frac{16}{3}b^mB_m+8D~,
\end{gather*} 
where the vertical bars denote the leading field components of the superfields.

It is straightforward (but tedious) to calculate the bosonic part of the Lagrangian in terms of the superfield components in Einstein frame, after elimination of the auxiliary fields and Weyl rescaling. We find
\begin{multline} \label{dhlag2}
e^{-1}\mathcal{L}=-\frac{1}{2}R-K_{ij^*}\partial^m A_i\partial_m\bar{A}_j-\frac{1}{4}F_{mn}F^{mn}-2J_{h\bar{h}}\partial_mh\partial^m\bar{h}-\frac{1}{2}J_{V^2}B_mB^m\\+iB_m(J_{Vh}\partial^mh-J_{V\bar{h}}\partial^m\bar{h})-\mathcal{V}~,
\end{multline}
where $h$, $\bar{h}$ are the Higgs field and its conjugate. We use the notation $J_{h\bar{h}}\equiv \frac{\partial^2J}{\partial h\partial\bar{h}}|$, $J_{Vh}\equiv \frac{\partial^2J}{\partial h\partial V}|$ and $J_{V^2}\equiv \frac{\partial^2J}{\partial V^2}|$. 
As regards the scalar potential, we get
\begin{equation} \label{finpot}
\mathcal{V}=\frac{1}{2}J_{V}^2+e^{K+2J}\left\{ (K+2J)^{IJ^*}(W_I+(K+2J)_IW)(\overbar{W}_{J^*}+(K+2J)_{J^*}\overbar{W})
-3W\overbar{W}\right\}~,
\end{equation}
where the capital Latin indices $I,J$ collectively denote all chiral superfields (as well as their lowest field components) including the Higgs superfield.

The standard $U(1)$ Higgs mechanism setting appears after employing the canonical function $J=\frac{1}{2}he^{2V}\bar{h}$.  As regards the Higgs sector, it leads to 
\begin{equation}
e^{-1}\mathcal{L}_{Higgs}=-\partial_mh\partial^m\bar{h}+iB_m(\bar{h}\partial^mh-h\partial^m\bar{h})-h\bar{h}B_mB^m-\mathcal{V}~.
\end{equation}
When parameterizing $h$ and $\bar{h}$ as
\begin{equation}
h=\frac{1}{\sqrt{2}}(\rho+\nu)e^{i\zeta},\;\;\;\bar{h}=\frac{1}{\sqrt{2}}(\rho+\nu)e^{-i\zeta}~,\label{paramh}
\end{equation}
where $\rho$ is the (real) Higgs boson, $\nu\equiv \langle h\rangle=\langle \bar{h}\rangle$ is the Higgs VEV, and $\zeta$ is the Goldstone boson, in the unitary gauge of $h\rightarrow h'=e^{-i\zeta}h$ and $B_m\rightarrow B'_m=B_m+\partial_m\zeta$, we reproduce the standard result \cite{Weinberg:1973ew}
\begin{equation}
e^{-1}\mathcal{L}_{Higgs}=-\frac{1}{2}\partial_m\rho\partial^m\rho-\frac{1}{2}(\rho+\nu)^2B_mB^m-\mathcal{V}~.
\end{equation}

The same result is also achieved by considering the super-Higgs mechanism where, in order to get rid of the Goldstone mode, we employ the super-gauge transformations \eqref{supergh} and \eqref{supergv}, and define the relevant field components of $Z$ and $i(Z-\overbar{Z})$ as
\begin{equation}
Z|=\zeta+i\xi~,\;\;\; \frac{i}{2}\overbar{\mathcal{D}}_{\dot{\alpha}}\mathcal{D}_{\alpha}(Z-\overbar{Z})|=\sigma_{\alpha\dot{\alpha}}^m\partial_m\zeta~.
\end{equation}
Examining the lowest components of the transformation \eqref{supergh}, we find that the real part of $Z|$ and $\overbar{Z}|$ cancels the Goldstone mode of \eqref{paramh}. Similarly, applying the derivatives $\overbar{\mathcal{D}}_{\dot{\alpha}}$ and  $\mathcal{D}_{\alpha}$ to \eqref{supergv} and taking their lowest components (recalling then $\overbar{\mathcal{D}}_{\dot{\alpha}}\mathcal{D}_{\alpha}V|=\sigma^m_{\alpha\dot{\alpha}}B_m$), we conclude that the vector field "eats up" the Goldstone mode indeed, as
\begin{equation}
B'_m=B_m+\partial_m\zeta~.
\end{equation}

\section{Adding a cosmological constant}

A cosmological constant (or dark energy) can be introduced into our framework without breaking any symmetries, via a simple modification of the Polonyi sector and its parameters $\alpha$ and $\beta$ introduced in Sec.~2.~\footnote{A similar idea was used in Ref.~\cite{Linde:2016bcz}, though in the different context, where the Polonyi potential was needed to prevent the real part of the stabilizer  field from vanishing at the minimum by imposing the condition $m_{gravitino}\ll m_{inflaton}$. In our approach, there is no stabilizer field, while the inflation comes from the D-type potential.}

Just adding a (very) small positive constant $\delta$ and assuming that $J'=0$ at the minimum of the potential modify the 
 (Minkowski) vacuum condition $V=0$ of Sec.~2 to
\begin{equation}
V=\mu^2e^{\alpha^2}\delta=m^2_{3/2}\delta~.\label{dsvac}
\end{equation}
By comparing the condition \eqref{dsvac} to Eq.~\eqref{vac1} we find a relation
\begin{equation}
(1+\alpha\beta+\alpha^2)^2-3(\alpha+\beta)^2=\delta~.\label{ccvac}
\end{equation}

A solution to Eqs.~\eqref{ccvac} and \eqref{vac2} with $V=m_{3/2}^2\delta$ is the true minimum, and it reads
\begin{equation}
\alpha=(\sqrt{3}-1)+\frac{3-2\sqrt{3}}{3(\sqrt{3}-1)}\delta+\mathcal{O}(\delta^2)~,\quad \beta=(2-\sqrt{3})+\frac{\sqrt{3}-3}{6(\sqrt{3}-1)}\delta+\mathcal{O}(\delta^2)~.
\end{equation}
This yields a {\it de Sitter} vacuum with the spontaneously broken SUSY after inflation.

Inserting the solution into the superpotential  and ignoring the $\mathcal{O}(\delta^2)$-terms, we find 
\begin{equation}
\langle \mathcal{W}\rangle=\mu(\alpha+\beta)=\mu(a+b-\frac{1}{2}\delta)~,
\end{equation}
where $a\equiv(\sqrt{3}-1)$ and $b\equiv(2-\sqrt{3})$ are the SUSY breaking vacuum solutions to the Polonyi parameters in the absence of a cosmological constant (Sec.~2).

\section{Scalar potential and vacuum stability}

For completeness, stability of our vacuum solutions should also be examined. On the one hand, in our model the vacuum stability is almost guaranteed because both functions ${J'}^2$ and $J''$ enter the scalar potential
\begin{equation}\label{spots}
V=\frac{1}{2}{J'}^2+\mu^2e^{A\bar{A}+2J}\biggl[
	|1+A\beta+A\bar{A}|^2-\bigg(3-2\fracmm{{J'}^2}{J''}\bigg)|A+\beta|^2
	\biggl]~
\end{equation}
with the positive sign, while the function $J''$ is required to be positive for the ghost-freedom. On the other hand, the only term with the negative sign in the scalar potential (\ref{spots})  is $-3|A+\beta|^2$ but it grows slower than the positive quartic term 
$|1+A\beta+A\bar{A}|^4$.

\begin{figure}
\centering
  \includegraphics[scale=.7]{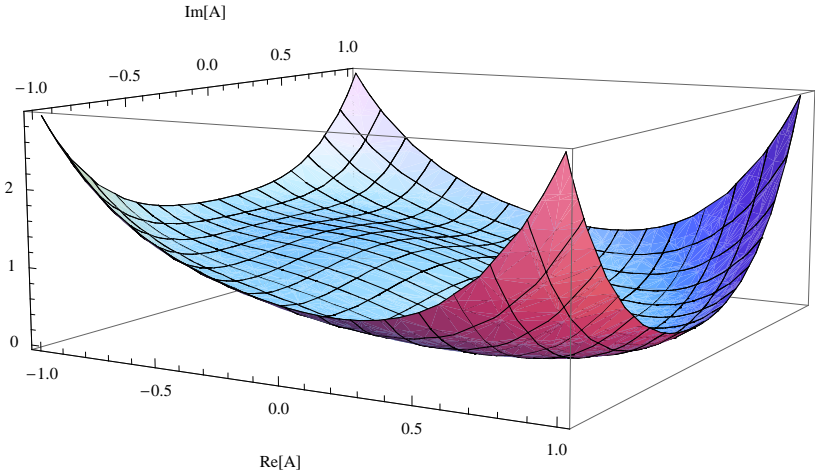}
  \caption{The scalar potential $\tilde{V}=\mu^{-2}e^{-A\bar{A}-2J}V$ as a function of $\text{Re}(A)$ and 
  $\text{Im}(A)$ at $J'=0$.}
  \label{f1}
\end{figure}

\begin{figure}
\centering
  \includegraphics[scale=.6]{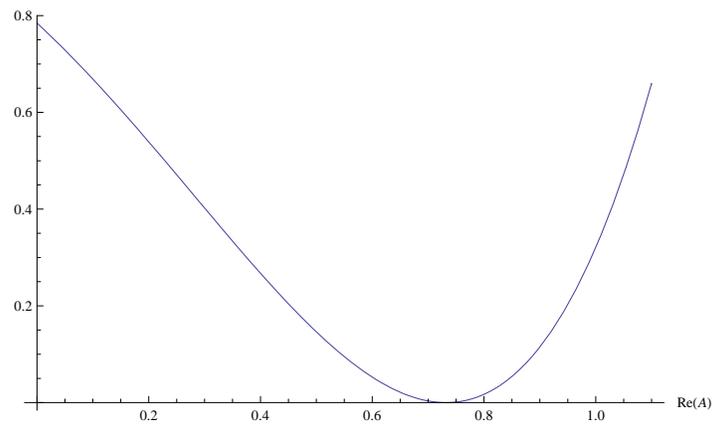}
  \caption{The real slice at $\text{Im}(A)=0$ of Fig.~\ref{f1} around the minimum of $\tilde{V}$.}
  \label{f2}
\end{figure}

The non-negativity of the scalar potential (\ref{spots}) for $|A|<1$ is not as apparent as that for $|A|\geq 1$. That is why we supply Figs.~\ref{f1} and \ref{f2} where the non-negativity becomes apparent too. In accordance to the previous Sec.~4, we can also add a positive cosmological constant that shifts the minimum to $V=m_{3/2}^2\delta$ describing a de Sitter vacuum.

\section{Conclusion}

Our new results are given in Secs.~3, 4 and 5. The new gauge-invariant formulation of our models can be used for unification of inflation with super-GUT in the context of supergravity, and has a single inflaton scalar field having a positive
definite scalar potential, a spontaneous SUSY breaking and a de Sitter vacuum after inflation. Our approach does not
preserve the $R$-symmetry.

Our upgrade of the earlier resuls in Ref.~\cite{Aldabergenov:2016dcu} is not limited to the generalized matter couplings in supergravity, given by Eqs.~(\ref{dhlag2}) and (\ref{finpot}). The standard approach to inflation in supergravity is based on the assumption that inflaton belongs to a chiral (scalar) multiplet. It leads to the well known problems such as the so-called 
$\eta$-problem, stabilization of other scalars,  getting SUSY breaking and a dS vacuum after inflation, etc. Though some solutions to these problems exist in the literature, they are rather complicated and include the additional "hand-made" input such as extra (stabilizing) matter superfields, extra (shift) symmetries or extra (nilpotency) conditions. We advocate another approach where inflaton is assumed to  belong to a massive vector multiplet, while SUSY breaking and a dS vacuum are achieved with the help of a Polonyi superfield. It is much simpler and more flexible than the standard approach.

Physical applications of our approach to super-GUT and reheating are crucially dependent upon the way how the fields present in our models  interact with the super-GUT fields. Consistency of sGUT with inflation may lead to some new constraints on both. For instance, inflaton
couplings to other matter have to be smaller than $10^{-3}$, in order to preserve flatness of the inflaton scalar potential and match the
observed spectrum of CMB density perturbations. In particular, Yukawa couplings of inflaton to right-handed (sterile) neutrino are crucial
to address the leptogenesis via inflaton decay and the subsequent reheating via decays of the right-handed neutrino into visible particles of the Standard Model.  Unfortunately, all this appears to be highly model-dependent at present. A derivation of
our supergravity models from superstrings, if any, is desirable because it would simultaneously fix those (unknown) interactions and
thus provide specific tools for a computation of reheating temperature, matter abundance, etc. after inflation, together with the low-energy predictions via gravity- or gauge- mediated SUSY breaking to the electro-weak scale --- see e.g., Ref.~\cite{Ellis:2016spb} for the previous studies along these lines. 
  
Our models can be further extended in the gauge-sector to the Born-Infeld-type gauge theory coupled to supergravity and other matter, along the lines of Refs.~\cite{Abe:2015fha,Aoki:2016tod}, thus providing further support towards their possible origin
in superstring (flux-)compactification.

\section*{Acknowledgements}

Y.A. is supported by a scholarship from the Ministry of Education, Culture, Sports, Science and Technology (MEXT) in Japan.
S.V.K. is supported by a Grant-in-Aid of the Japanese Society for Promotion of Science (JSPS) under No.~26400252, a TMU President Grant of Tokyo Metropolitan University in Japan, the World Premier International Research Center Initiative (WPI Initiative), MEXT, Japan, and the Competitiveness Enhancement Program of Tomsk Polytechnic University in Russia. The authors are grateful to the referees for their critical comments.

\bibliographystyle{utphys} 
%\bibliography{bibliography.bib}

\begin{thebibliography}{10}

\bibitem{Ade:2015xua}
{\bfseries Planck} Collaboration, P.~A.~R. Ade {\em et~al.}, ``{Planck 2015
  results. XIII. Cosmological parameters},''
\href{http://arxiv.org/abs/1502.01589}{{\ttfamily arXiv:1502.01589
  [astro-ph.CO]}}.
%%CITATION = ARXIV:1502.01589;%%.

\bibitem{Ade:2015lrj}
{\bfseries Planck} Collaboration, P.~A.~R. Ade {\em et~al.}, ``{Planck 2015
  results. XX. Constraints on inflation},''
\href{http://arxiv.org/abs/1502.02114}{{\ttfamily arXiv:1502.02114
  [astro-ph.CO]}}.
%%CITATION = ARXIV:1502.02114;%%.

\bibitem{Array:2015xqh}
{\bfseries BICEP2, Keck Array} Collaboration, P.~A.~R. Ade {\em et~al.},
  ``{Improved Constraints on Cosmology and Foregrounds from BICEP2 and Keck
  Array Cosmic Microwave Background Data with Inclusion of 95 GHz Band},''
  \href{http://dx.doi.org/10.1103/PhysRevLett.116.031302}{{\em Phys. Rev.
  Lett.} {\bfseries 116} (2016) 031302},
\href{http://arxiv.org/abs/1510.09217}{{\ttfamily arXiv:1510.09217
  [astro-ph.CO]}}.
%%CITATION = ARXIV:1510.09217;%%.

\bibitem{Yamaguchi:2011kg}
M.~Yamaguchi, ``{Supergravity based inflation models: a review},''
  \href{http://dx.doi.org/10.1088/0264-9381/28/10/103001}{{\em Class. Quant.
  Grav.} {\bfseries 28} (2011) 103001},
\href{http://arxiv.org/abs/1101.2488}{{\ttfamily arXiv:1101.2488
  [astro-ph.CO]}}.
%%CITATION = ARXIV:1101.2488;%%.

\bibitem{Ketov:2012yz}
S.~V. Ketov, ``{Supergravity and Early Universe: the Meeting Point of Cosmology
  and High-Energy Physics},''
  \href{http://dx.doi.org/10.1142/S0217751X13300214}{{\em Int. J. Mod. Phys.}
  {\bfseries A28} (2013) 1330021},
\href{http://arxiv.org/abs/1201.2239}{{\ttfamily arXiv:1201.2239 [hep-th]}}.
%%CITATION = ARXIV:1201.2239;%%.

\bibitem{Ferrara:2013rsa}
S.~Ferrara, R.~Kallosh, A.~Linde, and M.~Porrati, ``{Minimal Supergravity
  Models of Inflation},''
  \href{http://dx.doi.org/10.1103/PhysRevD.88.085038}{{\em Phys. Rev.}
  {\bfseries D88} no.~8, (2013) 085038},
\href{http://arxiv.org/abs/1307.7696}{{\ttfamily arXiv:1307.7696 [hep-th]}}.
%%CITATION = ARXIV:1307.7696;%%.

\bibitem{VanProeyen:1979ks}
A.~Van~Proeyen, ``{Massive Vector Multiplets in Supergravity},''
\href{http://dx.doi.org/10.1016/0550-3213(80)90345-4}{{\em Nucl. Phys.}
  {\bfseries B162} (1980) 376}.
%%CITATION = NUPHA,B162,376;%%.

\bibitem{Aldabergenov:2016dcu}
  Y.~Aldabergenov and S.~V.~Ketov,
  ``SUSY breaking after inflation in supergravity with inflaton in a massive vector supermultiplet,''
{{\em Phys. Lett. B} {\bfseries 761} (2016) 115},
  %doi:10.1016/j.physletb.2016.08.016
\href{http://arxiv.org/abs/1607.05366}{{\ttfamily arXiv:1607.05366 [hep-th]}}.

\bibitem{Polonyi:1977pj}
J.~Polonyi,
``Generalization of the Massive Scalar Multiplet Coupling to the
  Supergravity'', Hungary Central Inst. Res. KFKI-77-93 (1977, REC. JUL 1978), 5 p. KFKI-77-93, unpublished.
%%CITATION = KFKI-77-93;%%.

\bibitem{Ketov:1996bm}
S.~V.~Ketov,
``Solitons, monopoles and duality: from Sine-Gordon to Seiberg-Witten'',
{{\em Fortsch. Phys.} {\bfseries 45} (1997) 237},
  {{\ttfamily arXiv:9611209 [hep-th]}}.
%%43 citations counted in INSPIRE as of 23 Jan 2017

\bibitem{Wess:1992cp}
J.~Wess and J.~Bagger, "Supersymmetry and supergravity", Princeton University Press, Princeton, NJ, 
\newblock
1992.
\newblock
%%CITATION = INSPIRE-350988;%%.

\bibitem{Weinberg:1973ew}
  S.~Weinberg,
  ``General Theory of Broken Local Symmetries,''
  {{\em Phys. Rev. D} {\bfseries 7} (1973) 1068}.
  %doi:10.1103/PhysRevD.7.1068
  %%CITATION = doi:10.1103/PhysRevD.7.1068;%%
  %152 citations counted in INSPIRE as of 19 Dec 2016
  
  \bibitem{Linde:2016bcz}
  A.~Linde,
  ``On inflation, cosmological constant, and SUSY breaking,''
  {{\em JCAP} {\bfseries 1611} (2016) 002},
  % doi:10.1088/1475-7516/2016/11/002
{{\ttfamily arXiv:1608.00119 [hep-th]}}.
  %%CITATION = doi:10.1088/1475-7516/2016/11/002;%%
  %3 citations counted in INSPIRE as of 22 Dec 2016

  \bibitem{Ellis:2016spb}
 J.~Ellis, H.-J.~He and Z.-Z.~Xianyu,
  ``Higgs inflation, reheating and gravitino production in no-scale supersymmetric GUTs'',
{{\em JCAP} {\bfseries 1808} (2016) 068},
%doi:10.1088/1475-7516/2016/08/068
{{\ttfamily arXiv:1606.02202 [hep-th]}}.

  
 \bibitem{Abe:2015fha}
H.~Abe, Y.~Sakamura, and Y.~Yamada, ``{Massive vector multiplet inflation with
  Dirac-Born-Infeld type action},''
  \href{http://dx.doi.org/10.1103/PhysRevD.91.125042}{{\em Phys. Rev.}
  {\bfseries D91} (2015) 125042},
\href{http://arxiv.org/abs/1505.02235}{{\ttfamily arXiv:1505.02235 [hep-th]}}.
%%CITATION = ARXIV:1505.02235;%%.


\bibitem{Aoki:2016tod}
  S.~Aoki and Y.~Yamada,
  ``More on DBI action in 4D $\mathcal{N}=1$ supergravity,''
  {{\ttfamily arXiv:1611.08426 [hep-th]}}.
  
  %%CITATION = ARXIV:1611.08426;%%

 
\end{thebibliography}

\providecommand{\href}[2]{#2}\begingroup\raggedright
\endgroup

\end{document}